# Systematic Review of Approaches to Improve Peer Assessment at Scale & Possible Research Questions

Manikandan Ravikiran

mravikiran3@gatech.edu

## 1 BACKGROUND & INTRODUCTION

Peer Assessment is a task of analysis and commenting on student's writing by peers, is core of all educational components both in campus and in MOOC's. However, with the sheer scale of MOOC's & its inherent personalized open ended learning, automatic grading and tools assisting grading at scale is highly important. Previously in assignment 1 we presented survey on tasks of post classification, knowledge tracing and ended with brief review on Peer Assessment (PA), with some initial problems. In this survey we shall continue review on PA from perspective of improving the review process itself. As such rest of this review focus on three facets of PA namely Auto grading and Peer Assessment Tools (we shall look only on how peer reviews/auto-grading is carried), strategies to handle Rogue Reviews, Peer Review Improvement using Natural Language Processing. Following the literature review we present two research problem by synthesising reviews of both assignment 1 and 2. The consolidated set of papers and resources so used are released in https://github.com/manikandan-ravikiran/cs6460-Survey-2.

## 2 LITERATURE REVIEW

The literature review is divided in to three parts. In section in 2.1 we present a survey on PA tools, followed by review of approaches to handle rogue reviews 2.2 and end with possible open research problem in **??**. The paper coverage stats are in 1.

### 2.1 Automatic Grading and Peer Assessment tools

Automated grading is essential for scaling up MOOCs and MOOC based Master's programs. This area has seen multiple works over the past 2 decades, resulting in the creation of a sub-domain on its own. The earliest works during our survey date back to 1980's (Larkey, 1998), using Bayesian independence classifiers and K Nearest-Neighbor classifiers to assign scores to manually-graded essays with a human agreement of 0.88. After this, there is a large body of works

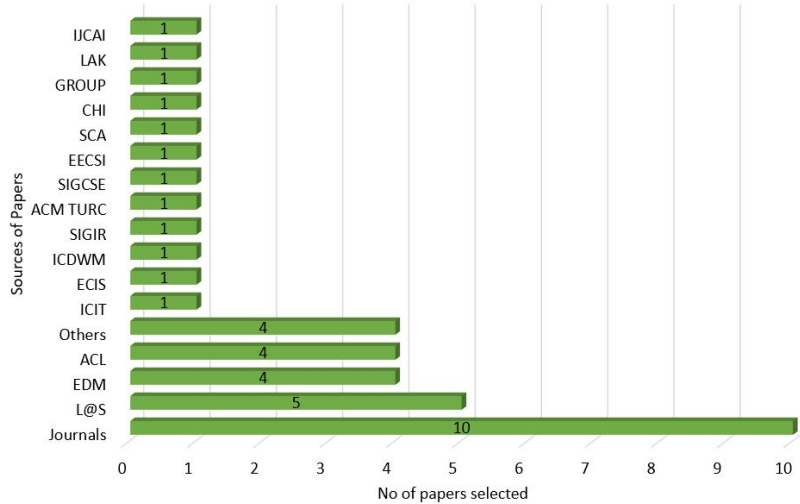

*Figure 1*—Coverage statistics of papers used in survey. ***Journals:=*** *Computer Education, Artificial Intelligence in Education, Intelligent Tutoring Systems, CBE life sciences education, Others:= Arxiv, Thesis*

diverging across multiple problems and solutions. The most notable recent work with a thorough analysis is by Geigle, Zhai, and Ferguson, (2016) which focuses on the systematic study of how to automate grading of a complex assignment using a medical case assessment as a test case. It proposes to solve this problem by using a supervised learning approach. The work introduces three general complementary types of feature representations and evaluates the feasibility and analyzes the results on 6 different rubrics to achieve accuracy in the range of 70% to 93% using supervised approaches. However much of this work focused on using classification for selecting structured relevant content from the medical case description to categories of questions, answers, quality, clarity, and application which is very specific to the medical domain and grading computation was done through ordinal regression on this work. As such the work, in general needs improvement for adaptation to other domains.

In a similar line, for non-textual domains we have EmbedInsight by (Piech et al., 2015), which is a grading tool for the course on embedded systems. The tool provides multiple experimental setups through a modular web services design with user interface and experimental setup back end. Further, the tool has modular testbeds for the execution of the designs in multiple sessions with visualization of latency and automatic grading of PWM signals. The work also presents prelim-inary results with a survey from students on their likeness score and also present the benefit of solubility based on submission as part of the university course. But the tool is very specific to embedded systems only.

On one side grading at scale research focuses on tools in new open domains or on complex assignment patterns. Alternatively, some works focus on building models to predict a student's future performance for a certain assessment activity within a single MOOC (Ren, Rangwala, and Johri, 2016). More specifically, this work performs real-time modeling by tracking the participation of a student within a MOOC and predicts the performance of a student on the next assessment within the course offering. The authors extend a multiple linear regression approach to the individual student and use multiple time levels, session-level, video, and homework related features. The work does report an average root mean squared error of 0.17 to show the benefits of the proposed idea. While this idea helps in modeling individual learner, the proposed approach is only usable for simplex assignments and sequential MCQ's.

On the other side, there are multifunctional peer assistance tools that cover every aspect involved in large scale peer assessment, notable among this is Peer studio (Kulkarni, Bernstein, and Klemmer, 2015), which presents PA and rapid feedback components with tools interoperability interface for other learning management systems. This tool takes a complementary approach compared to works that are described earlier. The tools main goal is to give formative feedback for work-in-progress. The work further shows that with fast and incremental feedback the grades of students are significantly higher and vice versa. The tool approaches problems of peer assistance in multiple parts, first being assisting in rubric selection, where the tool expects the reviewers to select the cell that most closely describes the submission. Following this tool uses a series of relevant words to scaffold the comments and requests for improvements, which is similar to (Geigle, Zhai, and Ferguson, 2016). Finally, for reviewer selection, the tool sends automatic emails with some implicit threshold logic. However, the tool has two major issues first of which where the authors do mention that most students did not create multiple drafts for revision, with the online class-topping over in-class version of the same course. Second, as we can see the rubrics are prior created and it expects the authors to read and provide comments to each, however, if we consider the complex assignment work described earlier Geigle, Zhai, and Ferguson, (2016) the tool is more disadvantageous. Yet this is the only tool (from my survey), that support rapid feedback.

Also, some tools focus on interactive feedback, notable among this is Critique-Kit (Ngoon et al., 2018) system that allows help reviewers give specific, actionable, and justified feedback. The tools show a guidance panel, that updates as progress in review happen. In the backend, there is a text classifier that categorizes the feedback as Specific, Actionable, Justified categories helping the reviewer. Also, it enables reusing feedbacks, where the suggestions from the tools update and adapt based on the prior explained categorization thereby giving ideas on how to improve the comments. The work further compares typical static suggestion and adaptive suggestions to find that people rarely used static suggestions and did not it help with adaptive suggestions were used more and found more helpful. There are multiple other peer assessment tools namely CTAS (Vogelsang and Ruppertz, 2015), ITPA (Lehmann and Leimeister, 2015), SWORD/Peerceptiv (Kaufman and Schunn, 2011) described earlier, Peer Scholar (Joordens, Desa, and Paré, 2013), Peer Grader (Gehringer, 2001). We shall revisit many other tools in Assignment 3.

## 2.2 Rogue Reviews

One of the biggest concerns voiced to date is Rogue Reviews (Staubitz et al., 2016), *which are comments that are insufficient, unusable by the peers either because of dishonesty, retaliation, competition or laziness* (Kulkarni, Bernstein, and Klemmer, 2015). Many of the tools that were described previously address rogue reviews including that of (Kulkarni, Bernstein, and Klemmer, 2015), (Kaufman and Schunn, 2011) use iterative analysis of comments to improve the overall feedback where the feedback is adapted to produce the best review possible. Instead of focusing on the tools (we shall return on this in assignment 3) here we will see alternate strategies that are explored to reduce rogue reviews in general. Many of these tools use these strategies as part of the process.

More specifically rogue reviews could be solved different perspectives based on review comments or review scores. In this work, we consider the following major approaches.

1. Improving numerical score given to assignment (See section 2.2.1)
2. Improving assessment of reviewers (score or/and review comments) by forming better groups (See section 2.2.2)
3. Improving review comments using natural language processing (See section

2.2.3)

*2.2.1 Grading Accuracy Improvement*

Grading accuracy/score improvement focuses on approaches to improve the accuracy of peer review scores to a relatively acceptable standard (Staubitz et al., 2016). Multiple works are done to date in both online and classroom setting, few of them are presented here. Score improvement typically focuses on various approaches to merge scores of different reviewers to find the correct score.

**Mean Aggregation and Hamer's algorithm:** Notable and most recent works in this line is by (Reily, Finnerty, and Terveen, 2009) which focuses on among many things, the accuracy of peer reviews. For this the work explores this by first creating a gold standard TA reviewed assignments and then collected 378 individual peer reviews for the six programming assignments. This was further analyzed & review for accuracy under several aggregation schemes. In general, the works find that students are harsher in terms of review compared to TA's, more specifically the work finds that On average, student review scores were 2.6 points (out of 100) lower than the TA's score. Further, the work focuses on improving the score through aggregation strategies, where it finds that taking mean aggregation of the scores by 3 reviewers improved results with a mean difference of 0.750. Additionally, the work also employs a second strategy to counter laziness, etc. through *smart aggregation*, where the work uses Hamer's algorithm for automatically calibrating peer review scores. The improvement is minimal compared to the mean approach however, we can see from that work that using Hamer's algorithm provides additional information about the quality of student reviews that can be used to provide an incentive for students to write good reviews. While the work, provides detailed evaluations it also has two major problems first being it is only applied to programming assignment where the scoring rubric is very easy and discrete compared to textual assignments. Second, the aggregation approach is very naive, multiple reviewer parameters could be used to improve the overall score aggregation, for example, time spent for review (less time spent means laziness or rogue reviews), etc. However, this requires details of clickstream information.

**Graphical Modeling:** Alternatively there are works, which employ more advanced techniques rather than simple mean aggregation. Famous work in this line is by (, 2013). The work formulated and evaluated a probabilistic peer grad-ing graphical model (PGM) for estimation of submission grades as well as grader bias and reliability. The work achieved this by creating one of the largest peer grading networks with 63k peer grade submissions. To begin with, the work uses three important data components namely grader bias, reliability, true grade, and observed grade. The work then developed three separate PGM's focusing on each of the aspects. Using these models the peer grading accuracy is calculated by simulating the expected score. The evaluation of the model is done in two steps. In step 1 the bias is estimated by running the model on all data with ground truth grades. In step 2, the authors sample random pool of peer grades with ground truth and use assessments so submitted to estimate the possible grade and calculate residual on the ground truth.

The authors compare the results on multiple benchmarks to show that probabilistic graphical model-based aggregation helps in reducing RMS error by 33%. The authors further identify biases by graders who do not spend adequate time on grading. While the paper shows multiple benefits in this form of aggregation and how it improves grade accuracy, there are no details on granular results across multiple courses, the impact of one domain course on another, grader selection bias (only says there could be selection bias). Multiple other works use PGM's models for improving the accuracy of grades notable and recent ones include (He, Hu, and Sun, 2019) which use Markov Chain Monte Carlo approaches and (Wu et al., 2015) which uses a Fuzzy Cognitive Diagnosis Framework (Fuzzy CDF).

**Content based re-ranking:** Alternative to PGM's and Aggregation, improving grading accuracy also focuses on exploiting contents in the assignments itself in the form of ranking problem (Luaces et al., 2015). The work approaches the problem of grading as the ranking of the assignment first by learning a scoring function of a sparse matrix of reviewers (N) v/s assignments (M), with few reviewers assigned to few assignments. This scoring function so learned on the sparse matrix is used to convert it into a dense matrix of scores. In a sense, the approach tries to predict scores for all the assignments from all the reviewers by modeling a function. These scores are then aggregated to obtain the final ranking. The authors approach the problem from a perspective of lack of availability of a large number of reviews, yet avoid the subjectivity of grader through preference learning (Luaces et al., 2015). The work compares the ranking so predicted using the Area under Curve metric against a baseline algorithm on multiple university course datasets. Finally, the authors show a comparison of results produced by the ranking approach against ground truth from 3 staff instructors of universities. The results from the paper also show that it is possible to produce good ranking despite sparse reviews and thereby helping peer assessment grading using content-based approaches. However, the approach does suffer from multiple open problems. Beginning with the approach of dealing with sparsity. Since the model uses sparse data to create dense metric one could argue that the matrix has very similar grades across the assignment, while the paper doesn't report anything on change in standard deviation or means of grades, based on results one could argue that the results are valid only for this datasets or dataset where grades range in very limited. Many complex assignments present a wide grading range where the proposed approach typically fails. Using contents to improve peer grading is a huge sub-domain in itself, during my survey I found many active groups working in this area. Some interesting ones which I did a review but didn't present in this document includes (Gütl, 2008), (Wang, Chang, and Li, 2008) and (Thomas et al., 2004).

*2.2.2 Better Reviewer Identification & Group Creation*

Typically in large scale MOOCs, reviewers are very diverse in terms of their knowledge and background, which in turn has an impact on the grading process. This is despite both the reviewer and author of the work is in the same course. In such a situation, it's better to pool authors and reviewers based on the same background resulting in better grading and improvement in the overall process. A wide variety of approaches have been studied in this angle. As such in this section, we review some of those works. In much of the literature, one will find this work reported as the task of group formation or group clustering, etc. Moreover, the literature is not only restricted to grouping for the sake of peer reviews but also other activities as part of MOOC as well. We shall first review the general Ed-tech perspective, then show more specific works on MOOC.

**Group Creation in General Edtech:** Interesting work in this line starts with (Ardaiz-Villanueva et al., 2011) which aimed to improve creativity with a focus on generating novel ideas in a university. More specifically, the work proposed a hypothesis that teams formed by subjects with high indexes of creativity and high indexes of affinity between them would obtain better results in originality and academic achievement than those teams with low indexes of creativity and high indexes of affinity, high indexes of creativity and low indexes of affinity and low indexes of creativity and low indexes of affinity (Ardaiz-Villanueva et al., 2011). To do this the work proposes idea of using a scoring mechanism called "creativity score", based on response length, number of responses provided by the 34 students from the Public University of Navarre Software Engineering Degree Program divided into eleven project groups each of which had three members except one with four similar to CS6460 course where ideas gathered in a brainstorm, combined with the creativity value, to suggest groups. The task of response gathering was carried followed by a series of interviews on a questionnaire. The work showed that high creative-high affinity team did produce extremely good results. However much of the is work focused on multiple aspects of the project including reviews. However, the data sample is very small compared to the scale of MOOC reviews, which is a big drawback. A more granular presentation of results was warranted.

Alternatively, there are works of Ounnas, (2010) which presents two similarity coefficients between users and learning objects and concentrates on automatic creation of properly matching collaborating groups based on an algorithmic approach. They are called a resemblance coefficient and relevance coefficients. The author proposes forming groups by reasoning on the semantics of data features generated by the participants, following this a ranking is made based on group constraints and strengths of created groups.

The data features generated include group formation are gender, nationality, age, previous marks, team role, and learning style. The work shows that with the proposed coefficients, features and parametric clustering algorithm it is possible to build a learning community group with its members towards a common objective. Additionally, the work also explains the ability to form groups suitable for distance learning. While the work shows descent results on group formation, the work doesn't show any results on MOOC review formation as such, however, analysis reveals that the approach is equally applicable in MOOC reviews. The work does have certain drawback where the analysis used simple data features which do not correlate with the aspect of performance.

While forming groups and pooling reviewers for assignment in MOOC course is one angle, some works take a top-down formation where the course itself is constructed such that groups are optimized. In this line, the works of Pollalis and Mavrommatis, (2008) is contemporary. The work proposes a method for course construction such that collaboration is easily achievable with a locus on common educational goals. The paper creates a matching approach for collaboration groups with appropriate learning objects to form courses suitable for the group so created. Further, the system so proposed also tracks learner's knowledge and group members such that the task is well achieved. The idea while interesting potentially creates under performing groups. Also, the work while presents result from the angle of collaboration, there is no detail on under performance.

**Group Creating in MOOCs:** Coming to MOOC reviews there are relatively lesser works (more survey required for this to be validated), notable among these is the research of (Lynda et al., 2017) which proposes to address peer assessment based on a combination of profile-based clustering and peer grading with the treatment of results. More specifically, the clustering part aims to group learners based on the parameters stored on learners' modeling within the MOOCs. The clustering is done based on three parameters (Lynda et al., 2017) namely *i) cognitive state, that contains the knowledge acquired during the activities of closed, semi-open and open questions ii) number of certificates obtained in MOOCs already followed iii) The preferences informed by the learner through a form or inferred during his / her learning by simple feedback.* The preferences considered concern with verbal learning and visual learning. These are used to form the first cluster, the first cluster is further subject to refinement based on the number of copies that each learner will have to correct. Because homogeneous groups are not creative. This stage tries to focus on distributing all the learners of the homogeneous groups formed prior in groups of K (4 to 10) learners the most heterogeneous possible. The authors do think that assessment with heterogeneous peers will have a positive impact on the scores and feedback to be provided to learners. Then following this the scores aggregated with weights given to each reviewer. Finally, the authors find that this way clustering has more significant groups, further resulting in complementary feedback in all constituted heterogeneous groups.

Then there is work by Wang, Lin, and Sun, (2007) which presents a tool called DIANA that uses genetic algorithms to achieve fairness, equity, flexibility, and easy implementation. More specifically DIANA was designed to create groups that exhibit internal diversity and external balance with other groups. To do this the work focuses on finding similarities in terms of similar in terms of heterogeneity, which fits well with our system goals of fairness (in the form of groups having the same size), equity (assigning all students to their most suitable group), flexibility (allowing teachers to address single or multiple psychological variables), and heterogeneity (guaranteeing individual diversity for promoting intra-group interactions). Further, the work tests the developed grouping approach on which group type offers more positive subjective comments concerning group partners, group outcomes, and the cooperative learning process. The authors find that DIANA-assigned groups correctly completed a significantly larger percentage of tasks that were set by the authors including the process of commenting. However, the work has two major problems first being total sample size which is only 66 students. Also, a system such as DIANA may be less useful to teachers who know their students well enough to develop their strategies for creating successful small learning groups. However, this kind of idea would be useful during peer grading group formations. There are many other works in similar line with usage of automatic algorithms to form groups (Graf and Bekele, 2006), (Fahmi and Nurjanah, 2018), and (Bekele, 2006).

### 2.2.3 *Peer Review improvement using Natural Language Processing:*

Much of the feedback from peers is descriptive in nature. Hence a linguistic analysis of these will be useful in multiple ways ranging from a simplification of review to be useful for the students to assess the reviewer himself based on quality. This area has garnered a plethora of works over the last decade with predominant works focusing on SWoRD benchmark tools. Notable and recent work in using NLP focuses on assessing the reviewer himself (Xiong, Litman, and Schunn, 2010). This work first detects the criticism feedback, and then predict the helpfulness of the recognized criticism feedback and reviewer performance. The work creates a new schema for the classification of feedback into 3 categories such as criticism, praise, and summary. Additionally, the reviews are also annotated for an indication of problems and solutions to those problems by the reviewers. The authors apply data mining techniques to achieve Cohen's kappa ($\kappa$) of 0.69 indicating the difficulty of the problem.

While assessing reviewer and contribution of review is one side, on another side, there are works that extend this to gather better reviews, most recent among this is by (Nguyen, Xiong, and Litman, 2016) where the authors use natural language processing to classify the comments into *Solution, Problem-only, or Non-criticism*. While the author ignores that non-criticism part of the comments they scaffold peers in revising problem-only comments (Nguyen, Xiong, and Litman, 2016). This is implemented as part of the Instant-feedback SWoRD Review system, where the tool displays a message at the top suggesting that comments may need to be revised to include solutions, followed by buttons representing the 3 possible reviewer responses. The developed tool was tested on 9 high-school Advanced Placement (AP) classes to see that had 16 incorrect triggers, achieving a precision 0.88. An extension of this work is also tested in classrooms (Nguyen, Xiong, and Litman, 2016)

In addition to the identification of various parts of review comments, there are also works that focus on the analysis of the quality of the review. Very recent work in this line is by (Ramachandran, Gehringer, and Yadav, 2016) which analyses review quality by using review content type, review relevance, review's coverage of a submission, review tone, review volume and review plagiarism as the metrics. While the review content, relevance is similar to that of Xiong, Litman, and Schunn, (2010) and Nguyen, Xiong, and Litman, (2016), coverage of a submission by a review using an agglomerative clustering technique to group the submission's sentences into topic clusters. The authors obtained f-scores of 0.67, 0.66 for the first two tasks, with coverage task presenting a cluster correlation of 0.49. Also in the rest of the tasks, the authors present results ranging from 44%-55% accuracy.

There are many other works which do focus on various part of the peer feedback namely Machine Classification of Peer Comments in Physics into praise, criticism, problem detection, solution suggestion, summary, or off-task. (Cho, 2008), predicting the quality of peer feedback regarding argument diagrams at sentence level (Nguyen and Litman, 2014), tool that employs data visualization at multiple levels of granularity, and provides automated analytic support using clustering and natural language processing (Xiong et al., 2012), topic modeling and its analysis in context of peer reviews as part of RevExplore tool (Xiong and Litman, 2013).

## 2.3 Summary & Findings

In this work, we reviewed three major sub-problems in PA, namely PA tools, approaches to handle rogue reviews and peer review improvement using NLP.

Firstly, coming to PA tools and automatic grading, we saw multiple PA tools were each presented different approaches to automatic grading depending on subjects, the context within MOOC, comment scaffolding, rapid feedback, and interactive feedback. At the same time, we plan to revisit tools in assignment 3. We did see two major aspects from works on PA tools and Auto-Grading, Firstly many works focus only on a programming assignment. Also, a plurality of works even with interactive and rapid feedback requires the reviewers to go through and respond to rubrics. Finally, the tools so reviewed including PeerStudio focuses on reviewers defining the rubric and answering each of the rubrics. Also, much of these tools address rogue reviews using approaches from works described in section 2.2.

Second coming to fixing issues with reviews, first there grade re-scoring in section 2.2. Where we saw that using multiple scores and aggregation does improve results, in this line the works are of wide variety based on the approaches used for score aggregation. In some cases, the ranking approaches are also exploited. However, one could see that all the score fixing approaches could be applied only for a single score for entire assignments. In the case of complex assignments like the works of Geigle, Zhai, and Ferguson, (2016) the approaches couldn't be applied, as the rubrics are given for multiple aspects of the content in the assignment.

Third, NLP has been bread and butter of the Ed-tech community in multiple problems including review improvement. Here we saw review improvement either in the form of iterative improvement assistance (Kaufman and Schunn, 2011) or Rapid Feedback (Kulkarni, Bernstein, and Klemmer, 2015). Alternatively, some works understand the granularity of the review comment to improve the overall scores (Xiong, Litman, and Schunn, 2010) and reviewer performance.

Comparing the previous three we can see that complex assignment review is still a big challenge where the reviewers go through step by step and assign reviews for each metric. While on one side this approach assists in rapid/iterative feedback, It's also time-consuming, resulting in lack of interest (Staubitz et al., 2016). While the problems in this survey, are described only with MOOC based research in mind, one would argue that this is also true for MOOC based Master's programs that use expert feedback which typically has high volume and high resource-producing courses (Joyner, 2017).

## 3 REFERENCES

[1]   Ardaiz-Villanueva, O., Chacón, X. N., Artazcoz, O. B., Acedo Lizarraga, M. L. S. de, and Acedo Baquedano, M. T. S. de (2011). "Evaluation of computer tools for idea generation and team formation in project-based learning". In: *Computers Education* 56, pp. 700–711.